\documentclass[twocolumn,amsmath,amssymb,amsfonts,superscriptaddress,floatfix,showpacs,prd,aps]{revtex4}
\newcommand{\del}{\partial}
\newcommand{\Tr}{\mathrm{Tr}}

\newcommand{\dslash}{\partial\!\!\!/}

\newcommand{\vev}[1]{\langle{#1}\rangle}

\def\sl#1{\setbox0=\hbox{$#1$}               % set a box for #1
   \dimen0=\wd0                                 % and get its size
   \setbox1=\hbox{/} \dimen1=\wd1               % get size of /
   \ifdim\dimen0>\dimen1                        % #1 is biggerbigger
      \rlap{\hbox to \dimen0{\hfil/\hfil}}      % so center / in box
      #1                                        % and print #1
   \else                                        % / is bigger
      \rlap{\hbox to \dimen1{\hfil$#1$\hfil}}   % so center #1
      /                                         % and print /
   \fi}                                         %
\usepackage{graphicx,color,mathrsfs,flafter}

\let\oldmarginpar\marginpar
\renewcommand\marginpar[1]{\-\oldmarginpar[\raggedleft\footnotesize #1]%
{\raggedright\footnotesize #1}}
\begin{document}
\title{\bf Meson fluctuations and thermodynamics of the  Polyakov loop extended
\\
quark-meson model }
\author{V. Skokov}
\email[E-Mail:]{V.Skokov@gsi.de}
\affiliation{%
GSI Helmholtzzentrum f\"ur Schwerionenforschung, D-64291
Darmstadt, Germany}
\author{B.~Stoki\'c}
\affiliation{%
GSI Helmholtzzentrum f\"ur Schwerionenforschung, D-64291
Darmstadt, Germany}
\author{B.~Friman}
\affiliation{%
GSI Helmholtzzentrum f\"ur Schwerionenforschung, D-64291
Darmstadt, Germany}
\author{K.~Redlich}
\affiliation{%
Institute of Theoretical Physics, University of Wroclaw, PL--50204 Wroc\l aw, Poland}
\affiliation{%
Theory Division, CERN, CH-1211 Geneva 23, Switzerland}
%\email[E-Mail:]{}
%\affiliation{}

%\pacs{}

%\date{\today}

\begin{abstract}
Thermodynamics and the phase structure of the  Polyakov loop-extended two flavor chiral quark--meson
model (PQM) are explored.  The analysis of the PQM model is based  on the functional renormalization
group (FRG) method. An appropriate truncation of the effective action with
quarks  coupled to background gluonic fields is introduced. Within this scheme,
we derive the renormalization group
flow equation for the scale-dependent thermodynamic potential at finite
temperature and density in
the presence of a symmetry breaking external field. 
%We determine the phase
%structure of the PQM model and
%use the Taylor expansion coefficients of the pressure to locate the
%position of the critical end point  in the phase diagram.
The influence  of fluctuations and of the
background gluon field on the properties of net-quark number density
fluctuations and their
higher moments is explored. We study the dependence of the kurtosis of
quark number fluctuations on the pion mass and show
that, in the presence of a symmetry breaking term, the fluctuations
lead to a smoothing of observables near the crossover transition.
\end{abstract}

\maketitle
%%%%%%%%%%%%%%%%%%%%%%%%%%%%%%%
%  SECTION I introduction
%%%%%%%%%%%%%%%%%%%%%%%%%%%%%%%

\section{Introduction}
The phase diagram of strongly interacting matter at nonzero baryon density and high temperature has
been a subject of growing interest in recent years. Calculations done within the
Lattice Gauge
Theory (LGT) show a clear separation between the confined--hadronic  and deconfined,  quark--gluon
plasma phase at finite temperature \cite{LGT}. Quantum Chromodynamics (QCD)
exhibits both dynamical
chiral symmetry breaking and confinement at finite temperature and densities. However, since QCD
thermodynamics   at large baryon densities is still not accessible for
first principle LGT calculations,
many phenomenological models and effective theories have been developed.

The hadronic properties at
low energy as well as the nature of the chiral phase transition at finite temperature and densities
have been successfully described and explored in such effective models.
Also, the physics of color deconfinement and its relation to chiral symmetry braking has been
recently  studied  in terms of  effective chiral
models~\cite{Gocksch,Buballa:review,Meisinger,Fukushima,Mocsy,PNJL,CS,DLS,Megias,IK,Fukushima:strong,Schaefer:PQM}.
The  idea  to extend the existing chiral models, such as the Nambu--Jona--Lasinio (NJL) or the
chiral quark--meson model,   by introducing couplings of quarks to  uniform temporal background
gauge fields (Polyakov loops)  was an   important step forward  in these  studies
~\cite{Fukushima,Schaefer:PQM}.

It was shown that the Polyakov loop extended
Nambu--Jona--Lasinio (PNJL) \cite{PNJL}
or quark meson (PQM)\cite{Schaefer:PQM} models reproduce essential
properties of QCD
thermodynamics obtained in the LGT calculations already in the mean-field
approximation. However, to
correctly account for the critical behavior and scaling properties  near the
chiral phase transitions
one needs to go beyond the mean-field approximation and include quantum fluctuations and
non-perturbative dynamics. This can be accounted for by using the functional renormalization group
(FRG) \cite{Wetterich,Morris,Ellwanger,Berges:review}. Until now this
method was applied
in the  NJL and quark--meson model, where the FRG equation was formulated for
quarks  coupled to
meson fields~\cite{Jungnickel,Berges:epjc,Schaefer:npa,Berges:prd_qm,Tetradis1,
Schaefer:vac,Schaefer:2006ds}.

In this paper  we propose a truncation of the PQM   model which is
suitable for a
functional renormalization group  analysis. We introduce an additional coupling of the chiral
condensate to a gluonic background via Polyakov loops. In this way the Polyakov
loop dynamics is
represented by a corresponding background temporal gauge field. We thus use
the functional
renormalization group  approach in order to include
fluctuations of the meson fields, while
the Polyakov loop  is treated on
a mean-field level. Consequently, our calculation lacks an FRG description
of the fluctuations of the Polyakov
loops. However, already in this approximation  we find an important effect
of the interactions of quarks with the effective gluon fields on the
thermodynamics. For
comparison we also present results for the chiral quark--meson model (QM)
without Polyakov loops in the FRG
approach  in order to make the difference in the thermodynamics of the two
effective models more transparent and to emphasize the role of the gluonic
background field. We also compare with the
PQM and QM models in mean-field approximation.

In this study we use the Taylor expansion method  to describe the thermodynamics
at finite chemical
potential $\mu$. We discuss the influence  of fluctuations and the
background gluon field on
the properties of the net-quark number density susceptibilities.
%We
%determine the phase diagram and the position of the critical end point (CEP) in
%the PQM model, by exploring
%the dependence of the order parameter and the quark number susceptibility on
%thermal parameters. 
Results on  the properties of  generalized
susceptibilities of the net quark number
density  in the presence of mesonic  fluctuations  in a  gluonic
background are presented.  We also explore ways to reduce the dependence on
the ultraviolet cutoff in the FRG approach.

The paper is organized as follows: In Sec. II we introduce
the PQM model. In Sec. III we formulate the
renormalization group equation for the thermodynamic potential in the PQM
model. In Sec. IV we consider  the   mean-field  approximation.   In
Sec. V we discuss the critical properties and thermodynamics of the PQM
model in the presence of mesonic fluctuations.

%%%%%%%%%%%%%%%%%%%%%%%%%%%%%%%%%%%%%%%%%%%%%%%%%%%%%
%%%%%%%%%%%%%%%%%%%%              PQM SECTION
%%%%%%%%%%%%%%%%%%%%%%%%%%%%%%%%%%%%%%%%%%%%%%%%%%%%%
\section{The Polyakov-quark-meson model}\label{sec:pqm}

The chiral quark--meson model  is  used as an effective realization of the low--energy sector of the
QCD. However, because the local $SU(N_c)$ invariance of QCD is replaced by
a global symmetry, one
cannot describe deconfinement phenomena in this model. Recently, it
was argued that,
by connecting  the chiral quark--meson model with the  Polyakov loop potential,
the confining
properties of QCD can be approximately accounted for~\cite{Fukushima,
Fukushima:strong, Schaefer:PQM}. Consequently, the Polyakov--quark--meson
(PQM) model effectively combines both the chiral and confining properties
of QCD.

The Lagrangian of the  PQM model  reads
\begin{eqnarray}\label{eq:pqm_lagrangian}
  {\cal L} &=& \bar{q} \, \left[i\sl{D} - g (\sigma + i \gamma_5
  \vec \tau \vec \pi )\right]\,q
  +\frac 1 2 (\partial_\mu \sigma)^2+ \frac{ 1}{2}
  (\partial_\mu \vec \pi)^2
  \nonumber \\
  && \qquad - U(\sigma, \vec \pi )  -{\cal U}(\ell,\ell^{*})\ .
\end{eqnarray}
The coupling between
the effective
gluon field and the quarks is implemented through the covariant derivative
\begin{equation}
 D_{\mu}=\del_{\mu}-iA_{\mu},
\end{equation}
where $A_\mu=g\,A_\mu^a\,\lambda^a/2$. The spatial components of the gluon
field are neglected, i.e. $A_{\mu}=\delta_{\mu0}A_0$.
Moreover, ${\cal U}(\ell,\ell^{*})$ is the effective potential for the gluon
field expressed in terms of
the thermal expectation values of the color trace of the  Polyakov loop and its conjugate

\begin{equation}
\ell=\frac{1}{N_c}\vev{\Tr_c L(\vec{x})},\quad \ell^{*}=\frac{1}{N_c}\vev{\Tr_c
L^{\dagger}(\vec{x})},
\end{equation}
with
\begin{eqnarray}
   L(\vec x)={\mathcal P} \exp \left[ i \int_0^\beta d\tau A_4(\vec
x , \tau)
  \right]\,,
\end{eqnarray}
where ${\mathcal P}$ stands for path ordering, $\beta=1/T$ and $A_4=i\,A_0$.

The $O(4)$ representation of the meson fields is
$\phi=(\sigma,\vec{\pi})$ and the corresponding $SU(2)_L\otimes SU(2)_R$ chiral  representation is
given by $\sigma+i\vec{\tau}\cdot\vec{\pi}\gamma_5$. There are $N_f^2=4$
mesonic degrees
of freedom coupled to $N_f=2$ flavors of quarks.

The purely mesonic potential of the model,  $U(\sigma,\vec{\pi})$,  is defined as
\begin{equation}
U(\sigma,\vec{\pi})=\frac{\lambda}{4}\left(\sigma^2+\vec{\pi}
^2-v^2\right)^2-c\sigma,
\end{equation}
while the  effective potential of the gluon field is parametrized
to preserve the $Z(3)$ invariance:
\begin{equation}
 \frac{{\cal U}(\ell,\ell^{*})}{T^4}=
-\frac{b_2(T)}{2}\ell^{*}\ell
-\frac{b_3}{6}(\ell^3 + \ell^{*3})
+\frac{b_4}{4}(\ell^{*}\ell)^2\,\label{eff_potential}.
\end{equation}
The parameters,
\begin{eqnarray}
\hspace{-4ex}
  b_2(T) &=& a_0  + a_1 \left(\frac{T_0}{T}\right) + a_2
  \left(\frac{T_0}{T}\right)^2 + a_3 \left(\frac{T_0}{T}\right)^3\,
\end{eqnarray}
with  $a_0 = 6.75$, $a_1 = -1.95$, $a_2 = 2.625$, $a_3 = -7.44$, $b_3 = 0.75$ and $b_4 = 7.5$ were
chosen  to reproduce the equation of state of pure gauge degrees of freedom  calculated on the
lattice. At the temperature  $T_0=270$ MeV, the critical temperature obtained for pure
gauge theory, the potential~(\ref{eff_potential}) yields  a first order phase
transition.

%Thus, at low
%temperatures the  ${{\cal U}(\ell,\ell^{*})}$   have an absolute minimum at $\ell=0$ whereas for
%temperatures $T>T_0$  the minimum is shifted to a finite value of $\ell$.
%At asymptotic high temperatures  the Polyakov loop expectation value tends to unity, $\ell\to1$.

%The thermodynamic properties  of the PQM model are calculated  from the thermodynamic potential
%\begin{eqnarray}  \Omega &=& {\mathcal U}(\ell,\ell^{*}) +
%  U(\sigma) + \Omega_{\bar qq}(\ell,\ell^{*},\sigma).
%\end{eqnarray}
%which  is obtained from the mean field Lagrangian (\ref{eq:pqm_lagrangian}) by introducing averaged
%meson fields and  integrating the  fermions within the Nambu--Gor'kov formalism.

%%%%%%%%%%%%%%%%%%%%%%%%%%%%%%%%%%%%%%%%%%%%%%%%%%%%%
%%%%%%%%%%%%%%%%%%%%              FRG SECTION
%%%%%%%%%%%%%%%%%%%%%%%%%%%%%%%%%%%%%%%%%%%%%%%%%%%%%
\section{The FRG method in the  PQM model}\label{sec:rg}

%One way  to  go beyond the mean field approximation  and  to include quantum fluctuations is the
%approach based  on  the FRG method.
%In our studies, based on the Polyakov loop extended quark--meson
%model, the FRG flow equation has not been considered until now in the literature.
%In this  Section we formulate the FRG method
%to account fluctuations of mesonic fields. The Polyakov loop variable is to be
%treated on the mean field level.
%Then we apply the obtained thermodynamic potential to study the importance of quantum
%fluctuations in the  PQM model near phase transition. In this context,  we  will analyze the
%properties of the net quark number fluctuations, its higher moments  and kurtosis.

The functional renormalization group  is an important tool for addressing
nonperturbative problems within the quantum field
theory. It is based on an infrared (IR) regularization with the momentum
scale parameter, $k$, of
the full propagator which turns the corresponding effective action into a scale dependent functional
$\Gamma_k$~\cite{Wetterich, Morris, Ellwanger, Berges:review}.  The change of  $\Gamma_k$ with the
change of the momentum scale  is described through the flow equation:
\begin{eqnarray}\label{eq:FRG}
\del_k\Gamma_k[\Phi,\psi]&=&\frac{1}{2}\Tr\left\{\del_k R_{kB}\biggl(\Gamma_k^{(2,0)}
[\Phi,\psi]+R_{kB}\biggr)^{-1}\right\}\nonumber \\
&&\hspace*{-7mm}-\Tr\left\{\del_k R_{kF}\left(\Gamma_k^{(0,2)}[\Phi,\psi]+R_{kF}\right)^{-1}\right\},
\end{eqnarray}
where $\Gamma_k^{(2,0)}$ and  $\Gamma_k^{(0,2)}$  denote the second
functional derivative of $\Gamma_k[\Phi,\psi]$ with
respect to the bosonic ($\Phi$) and fermionic ($\psi$) fields,
respectively. These derivatives  correspond  to the inverse of the
full propagators at the scale $k$. 
The trace    in  Eq.~(\ref{eq:FRG}) denotes  a momentum integration and a summation
over all internal
indices (e.g. flavor, color, and/or Dirac). Here $\Phi$ and $\psi$ denote
bosonic and
fermionic fields, respectively. The effective average action, $\Gamma_k$, governs the dynamics of a
theory at a momentum scale $k$ and interpolates between the bare action, $\Gamma_{k=\Lambda}\equiv S$,
and the full quantum effective action, $\Gamma_{k=0}=\Gamma$. The regulator function, $R_k$, describes
how the small momentum modes are cut off and is to some extent
arbitrary~\cite{Berges:review}. The derivative of this
function, $\partial_k R_k$, implements the Wilsonian idea of successively
integrating out momentum shells.

In the PQM model the formulation of the  FRG flow equation  (\ref{eq:FRG})  would required an
implementation of the Polyakov loop as a dynamical field. However, in this
exploratory calculation we treat the Polyakov loop on the mean-field level. This
allows us to
formulate the FRG flow equation for the truncated effective PQM
action in Euclidean space-time ($t\to - i \tau$):
\begin{eqnarray}\label{eq:trunc}
\Gamma_k &=& \int d^4x\left\{\frac{1}{2}\left(\del_{\mu}\phi\right)^2 +
\bar{q}\dslash q   \right.\nonumber \\
&& +\left.
g\bar{q} \, (\sigma + i
  \vec \tau\cdot \vec \pi\gamma_5 )\, q + U_k(\rho)
  \right\},
\end{eqnarray}
where the  field $\rho$ is given by
\begin{equation}\label{eq5}
\rho=\frac{1}{2}\phi^2=\frac{1}{2}\left(\sigma^2+\vec{\pi}^2 \right).
\end{equation}
The finite quark chemical potential, $\mu$, and the gauge potential, $A_0$,
are introduced through the following substitution of the time derivative
\begin{equation}
\del_\tau\to\del_\tau-(\mu+iA_4).
\end{equation}

Since  the gauge potential, $A_0$,
is  treated as an effective background gluonic field, which
does not flow with the scale parameter $k$,
the explicit contribution of the Polyakov
loop potential to the truncated effective action is suppressed at this stage.
Later on we will restore the Polyakov loop potential.
%(\ref{eff_potential}).

\begin{widetext}
As in the previous studies~\cite{SFR}, we employ the optimized regulator
functions~\cite{Litim:opt} in the numerical implementation.
%% rephrase?
For bosons, this  cutoff function depends  only on the spatial components of
the momentum,
\begin{equation}
R^{\mathrm{opt}}_{B,k}(\mathbf{q}^2)=(k^2-\mathbf{q}^2)\theta(k^2-\mathbf{q}^2),
\end{equation}
whereas for  fermions we use~\cite{SFR}
\begin{equation}
R^{\mathrm{opt}}_{F,k}(\mathbf{q})=\left(\sqrt\frac{(q_0+i\alpha_0)^2+k^2}{(q_0+i\alpha_0)^2+\mathbf{q}^2}-1\right)
 (\sl{q}+i\gamma^0\alpha_0)
\theta(k^2-\mathbf{q}^2),
\end{equation}
where $\alpha_0=\mu+iA_0$ is modified due to the coupling to the
background gluon field.

Using Eq.~(\ref{eq:trunc}), with the cutoff functions given above, and the
relation $\Omega=T\,\Gamma$ ~\cite{Wetterich2}, we obtain the
flow equation for the scale dependent grand canonical potential for
the quark and mesonic subsystems
\begin{eqnarray}\label{eq:frg_flow}
\del_k \Omega_k(\ell, \ell^*; T,\mu)&=&\frac{k^4}{12\pi^2}
 \left\{ \frac{3}{E_\pi} \Bigg[ 1+2n_B(E_\pi;T)\Bigg]
 +\frac{1}{E_\sigma} \Bigg[ 1+2n_B(E_\sigma;T)
    \Bigg]   \right. \\ \nonumber && \left. -\frac{4 N_c N_f}{E_q} \Bigg[ 1-
N(\ell,\ell^*;T,\mu)-\bar{N}(\ell,\ell^*;T,\mu)\Bigg] \right\}.
\end{eqnarray}
Here $n_B(E_{\pi,\sigma};T)$ is the bosonic  distribution function
\begin{equation*}
n_B(E_{\pi,\sigma};T)=\frac{1}{\exp({E_{\pi,\sigma}/T})-1}
\end{equation*}
with the pion and sigma energy
\begin{equation*}
E_\pi = \sqrt{k^2+\overline{\Omega}^{\,\prime}_k}\;~,~ E_\sigma
=\sqrt{k^2+\overline{\Omega}^{\,\prime}_k+2\rho\,\overline{\Omega}^{\,
\prime\prime} _k}.
\end{equation*}
where the primes denote derivatives with respect to $\rho$ and
$\overline{\Omega}=\Omega+c\sigma$.
The functions $N(\ell,\ell^*;T,\mu)$ and $\bar{N}(\ell,\ell^*;T,\mu)$, defined
by
\begin{eqnarray}\label{n1}
N(\ell,\ell^*;T,\mu)&=&\frac{1+2\ell^*\exp[\beta(E_q-\mu)]+\ell \exp[2\beta(E_q-\mu)]}{1+3\ell \exp[2\beta(E_q-\mu)]+
3\ell^*\exp[\beta(E_q-\mu)]+\exp[3\beta(E_q-\mu)]},  \\
\bar{N}(\ell,\ell^*;T,\mu)&=&N(\ell^*,\ell;T,-\mu),
\label{n2}
\end{eqnarray}
are fermionic distributions, modified due to the coupling to gluons, and
\begin{equation}
\label{dispertion}
E_q =\sqrt{k^2+2g^2\rho}
\end{equation}
is the quark energy. In the absence of
the background gluon field, e.g. in the QM model, $\ell,\ell^*\to 1$ and the
usual Fermi-Dirac distribution function is
recovered; $N(1,1;T,\mu)=n_F(E_q;T,\mu)=1/\{\exp[(E_q-\mu)/T]+1\}$.
\end{widetext}

In order to solve the flow equation (\ref{eq:frg_flow}), we expand
$\overline{\Omega}_k(T,\mu)$,
in a Taylor series around the scale-dependent minimum
at $\sigma_k=\sqrt{2\rho_k}$:
\begin{equation}\label{eq:taylor}
\overline{\Omega}_k=\sum_{i=0}^m \frac{a_{i,k} }{i!}(\rho-\rho_k)^i.
\end{equation}
As in Ref.~{\cite{SFR}}, we truncate the expansion at $m=3$.
The Taylor coefficients, $a_{i,k}$, are functions of the scale,
$k$,
the temperature, $T$, the chemical potential, $\mu$, and the Polyakov loop,
$\ell$ and $\ell^*$.

The minimum of the thermodynamic potential  is determined by the stationarity
condition
\begin{equation}
\left. \frac{d \Omega_k}{ d \sigma} \right|_{\sigma=\sigma_k}=\left. \frac{d
\overline{\Omega}_k}{ d \sigma} \right|_{\sigma=\sigma_k} - c =0.
\label{eom_sigma}
\end{equation}
%We also keep an  explicit symmetry
%breaking term $c$ that allows  to fix the  pion masses to their physical values.

The set of flow equations for each expansion coefficient, $a_{i,k}$,  arising from the Eq.~(\ref{eq:taylor}) is solved
numerically with an initial cutoff $\Lambda=1.2$ GeV. The initial
conditions, $a_{i,k=\Lambda}$ and $c$, are chosen
to  reproduce  the physical pion mass $m_{\pi}=138$ MeV, the pion decay constant
$f_{\pi}=93$ MeV, the sigma mass $m_{\sigma}=700$ MeV and the constituent quark mass $m_q=335$ MeV
at the scale $k=0$ in vacuum.
The symmetry breaking term, $c=m_\pi^2 f_\pi$, corresponds to an external
field and consequently does not flow. In this work, we neglect the flow of the
Yukawa coupling, $g$.

By solving equation  (\ref{eq:taylor})  we obtain a thermodynamic potential
for the quark and mesonic subsystems, $\Omega_{k\to0} (\ell, \ell^*;T, \mu)$,
as a function of Polyakov loop variables $\ell$ and $\ell^*$. So far these variables are
arbitrary. The { full thermodynamic potential  in the PQM model,
including quark, meson and
gluon degrees of freedom}
is obtained by adding  to  $\Omega_{k\to0} (\ell, \ell^*;T, \mu)$ the
effective gluon  potential ${\cal U}(\ell, \ell^*)$
from   Eq. (\ref{eff_potential}),
\begin{equation}
\Omega(\ell, \ell^*;T, \mu) = \Omega_{k\to0} (\ell, \ell^*;T, \mu) + {\cal U}(\ell, \ell^*).
\label{omega_final}
\end{equation}
At a given $T$ and
$\mu$,
the Polyakov loop variables, $\ell$ and $\ell^*$ are determined by the
stationarity conditions:
\begin{eqnarray}
\label{eom_for_PL_l}
&&\frac{ \partial   }{\partial \ell} \Omega(\ell, \ell^*;T, \mu)  =0, \\
&&\frac{ \partial   }{\partial \ell^*}  \Omega(\ell, \ell^*;T, \mu)   =0.
\label{eom_for_PL_ls}
\end{eqnarray}
{ Within the FRG approach, the thermodynamic potential  of the QM model is
obtained from Eq. (\ref{omega_final}),
by dropping the effective Polyakov
loop potential  and  setting $\ell=\ell^*=1$. }

{In the following we explore the  thermodynamics of the PQM-FRG model. The
role of the gluonic sector is assessed by juxtaposing the results
obtained within the PQM and QM models, while the effect of mesonic fluctuations
in the PQM model
are deduced  by comparing with the mean-field
approximation.}

% Fig 1
\begin{figure*}
\includegraphics*[width=7.5cm]{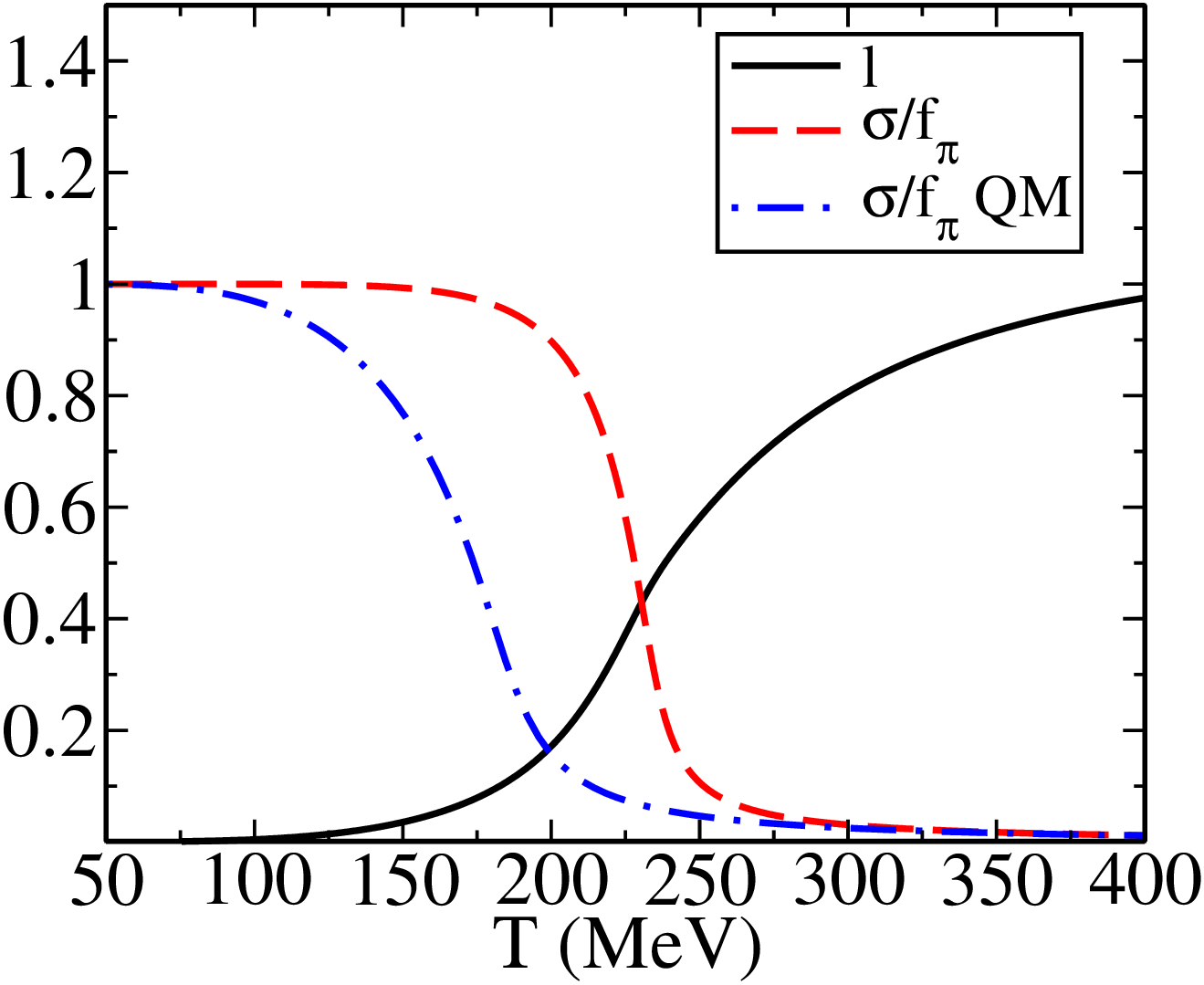}
\includegraphics*[width=7.5cm]{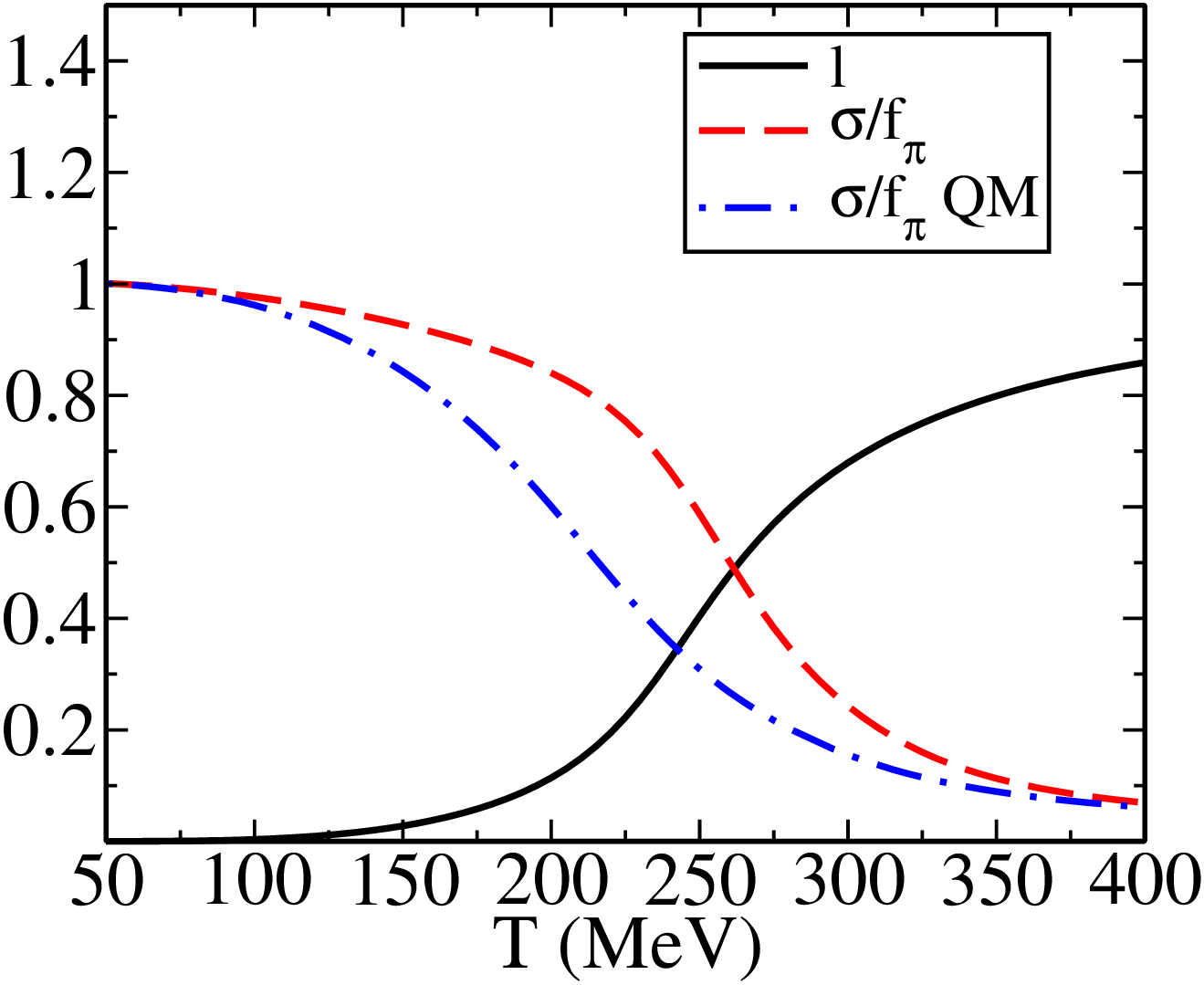}
\caption{Thermal average of the Polyakov loop, $\ell$, and of the order
parameter of the
chiral phase transition, $\langle \sigma\rangle$,  as functions of temperature
at zero baryon chemical potential
in the mean-field approximation(left panel) and  in  FRG approach
(right panel).
The solid and dashed lines are obtained in the PQM model, while the dash-dotted
lines correspond to the QM model.
}
\label{fig:fields}
\end{figure*}
% Fig 2
\begin{figure*}
\includegraphics*[width=7.5cm]{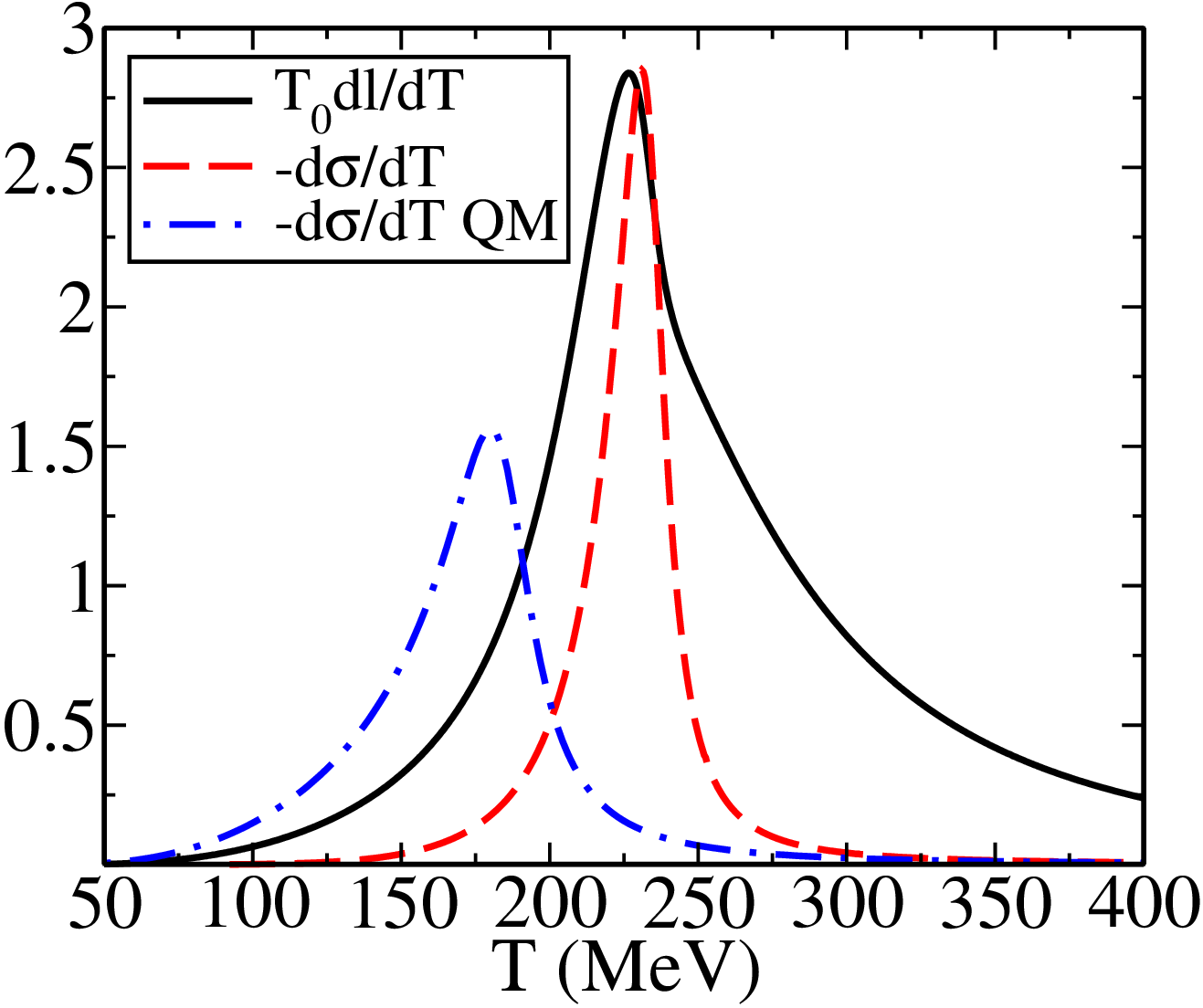}
\includegraphics*[width=7.5cm]{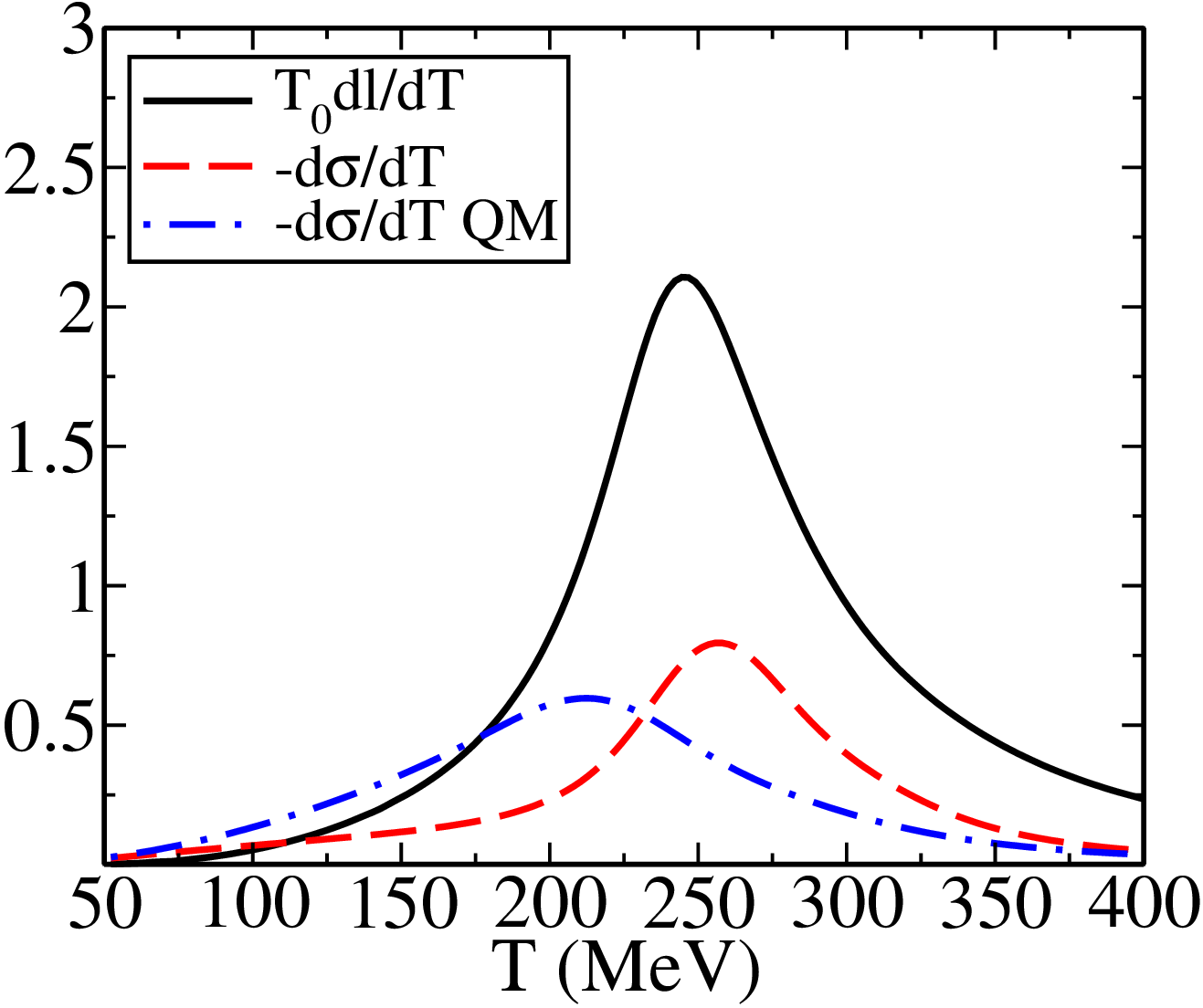}
\caption{Temperature derivatives of thermal average of the Polyakov loop,
$\ell$, and of the order parameter of the
chiral phase transition, $\langle \sigma\rangle$,  as functions of temperature
at zero baryon chemical potential in the mean-field approximation(left panel)
and  in  the FRG approach
(right panel). The notation is the same as in Fig.~\ref{fig:fields}.
}
\label{fig:derivatives}
\end{figure*}

\section{The mean-field approximation} \label{sec:mf}

The importance of mesonic fluctuations is best illustrated by comparing
the FRG approach with the mean-field (MF) approximation for mesons. In the latter,
both quantum and thermal fluctuations are neglected and the mesonic fields
are replaced by their classical expectation values. In the formulation
presented in the
previous section, the mean-field approximation is recovered by omitting
the pion- and sigma-meson contributions in  Eq. (\ref{eq:frg_flow}).
Alternatively, this approximation can be obtained directly
from the partition function for the Lagrangian (\ref{eq:pqm_lagrangian}) by
integrating out the
quark degrees of freedom. These approaches are equivalent up to a surface
term, which is negligible for large cuttoffs,  $\Lambda\to \infty$.
We explicitly retain the divergent vacuum contribution,
which
is regularized by an ultraviolet cuttoff $\Lambda_{MF}$ . The importance of
this
contribution was demonstrated in Refs.~\cite{Nakano:2009ps,MFonVT}.

\begin{widetext}
A detailed derivation of the mean-field approximation of  PMQ model
can be found in Ref.~\cite{Schaefer:2007pw} and references therein.
Here we present only the final result for thermodynamic potential
\begin{equation}
\Omega_{MF} = {\cal U}(\ell,\ell^*) + U(\langle\sigma\rangle, \langle\pi\rangle=0) + \Omega_{q\bar{q}} (\langle\sigma\rangle,\ell,\ell^*).
\label{Omega_MF}
\end{equation}
The contribution of quarks with mass $m_q=g\langle\sigma\rangle$ is given by
\begin{equation}
\Omega_{q\bar{q}} (\langle\sigma\rangle, \ell,\ell^*) = - 2 N_f T \int \frac{d^3 p}{(2\pi)^3} \left\{
 \frac{N_c E_q}{T} \theta( \Lambda_{MF}^2 - p^2)  + \ln
g^{(+)}(\langle\sigma\rangle, \ell, \ell^*; T, \mu) +  \ln
g^{(-)}(\langle\sigma\rangle,\ell, \ell^*; T, \mu) \right\},
\label{Omega_MF_q}
\end{equation}
where
\begin{eqnarray}
\label{g}
g^{(+)}(\langle\sigma\rangle,\ell, \ell^*; T, \mu) &=& 1 + 3 \ell
\exp[-(E_q-\mu)/T] + 3 \ell^*\exp[-2(E_q-\mu)/T] + \exp[-3(E_q-\mu)/T], \\
g^{(-)}(\langle\sigma\rangle,\ell, \ell^*; T, \mu) &=& g^{(+)} (\langle\sigma\rangle,\ell^*, \ell; T, -\mu);
\end{eqnarray}
and  $E_q = \sqrt{p^2+m_q^2}$ is the quark quasi-particle energy.
\end{widetext}
The equations of motion for the mean fields are
obtained by requiring that the thermodynamic potential be stationary with
respect to changes of $\sigma$, $\ell$ and $\ell^*$:
\begin{equation}
\frac{\partial \Omega_{MF}}{\partial \sigma} = \frac{\partial \Omega_{MF}}{\partial \ell} = \frac{\partial \Omega_{MF}}{\partial \ell^*} =0.
\label{EOM_MF}
\end{equation}

The model parameters are fixed to reproduce the same vacuum physics as in the
FRG calculation, as
described in the previous section. The additional free
parameter, $\Lambda_{MF}$,
is chosen so as to reproduce the quark condensate in vacuum
$\langle\bar{u}u\rangle=-$(260 MeV)$^3$; we find $\Lambda_{MF}=674$ MeV.

\section{Fluctuations and thermo\-dynamics of the PQM model}\label{sec:thermo}

The thermodynamic potential obtained in section  \ref{sec:rg}
can be used to explore
the influence  of
the gluonic background field on the thermodynamics in the PQM model
including the effect of
fluctuations.
In previous studies of the chiral quark--meson model  within the FRG approach,
it was
shown that non-perturbative meson contributions modify the position
of the chiral boundary
and the critical end point (CEP) in the $(T,\mu)$-plane. The
pseudocritical
temperature and chemical potential is usually identified by
a maximum in the temperature derivative of the order parameter
or in the chiral susceptibility.
In Figs.~\ref{fig:fields} and \ref{fig:derivatives} we show the dependence of
$\langle \sigma\rangle$
and Polyakov loop $\ell$ and of
their temperature derivatives computed in the FRG approach as well as in the
mean-field approximation for the PQM and QM models.

As expected,  for a physical pion mass the model exhibits a smooth crossover
as a function of temperature. Moreover, as shown in Fig.
\ref{fig:derivatives},
the temperature derivatives of both the Polyakov loop and the chiral order
parameter
exhibit peak structures at approximately the same temperature, $T_c=257$
MeV in
the { PQM-FRG} and $T_c=230$ MeV in the { PQM-MF} model.
In general the relation between the deconfinement and chiral phase
transitions is not well established. Recent LGT calculations
lead to conflicting
conclusions concerning the relative position of these transitions
\cite{LGT,Aoki:2006br}.

A comparison of the QM and PQM models shows that the inclusion of gluon
degrees of freedom
shifts the chiral phase transition to higher temperatures both with and
without mesonic fluctuations. The latter lead to a smoothing of
the transition, as shown in Figs.~\ref{fig:fields} and
\ref{fig:derivatives}.

\begin{figure*}
\includegraphics*[width=7.5cm]{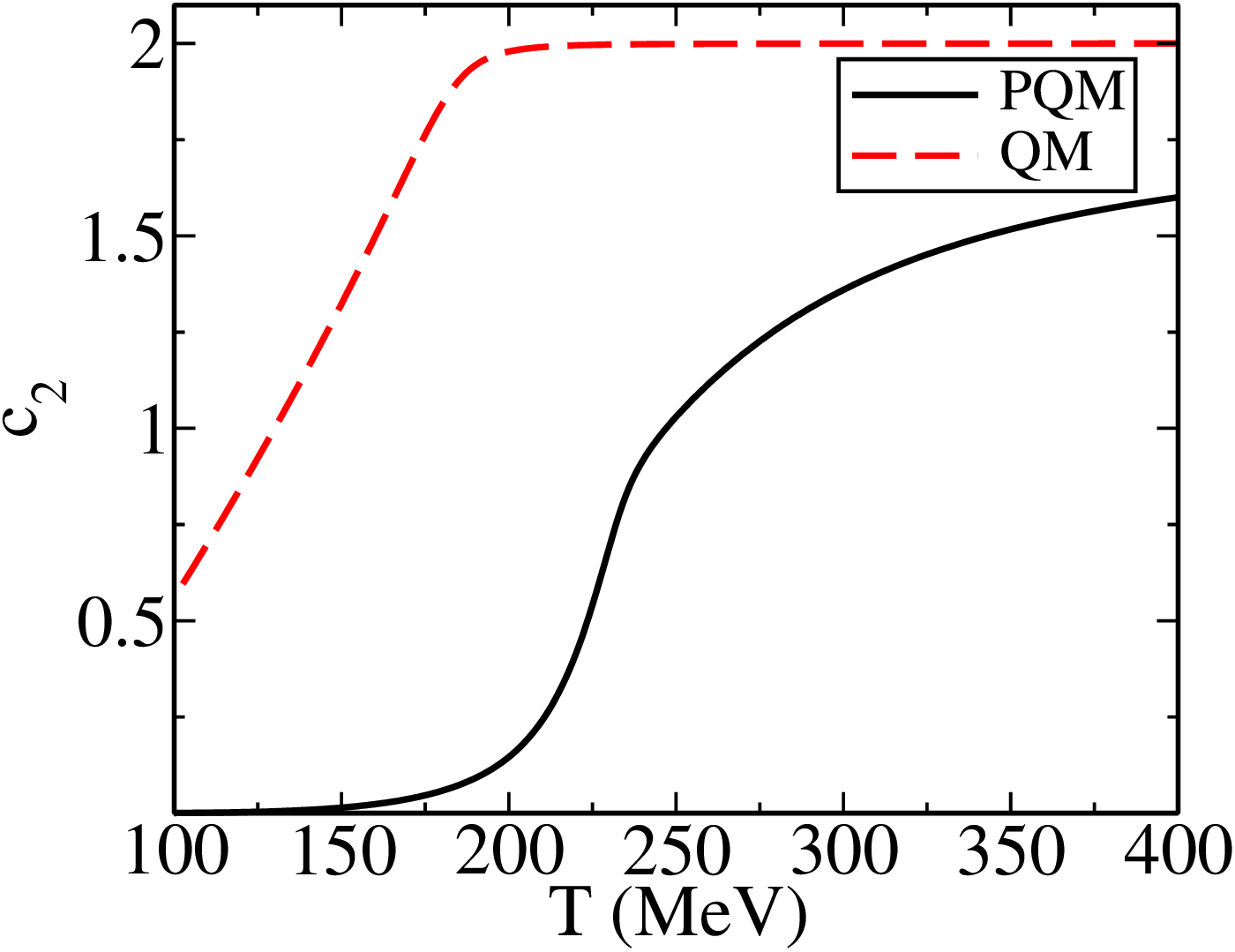}
\includegraphics*[width=7.5cm]{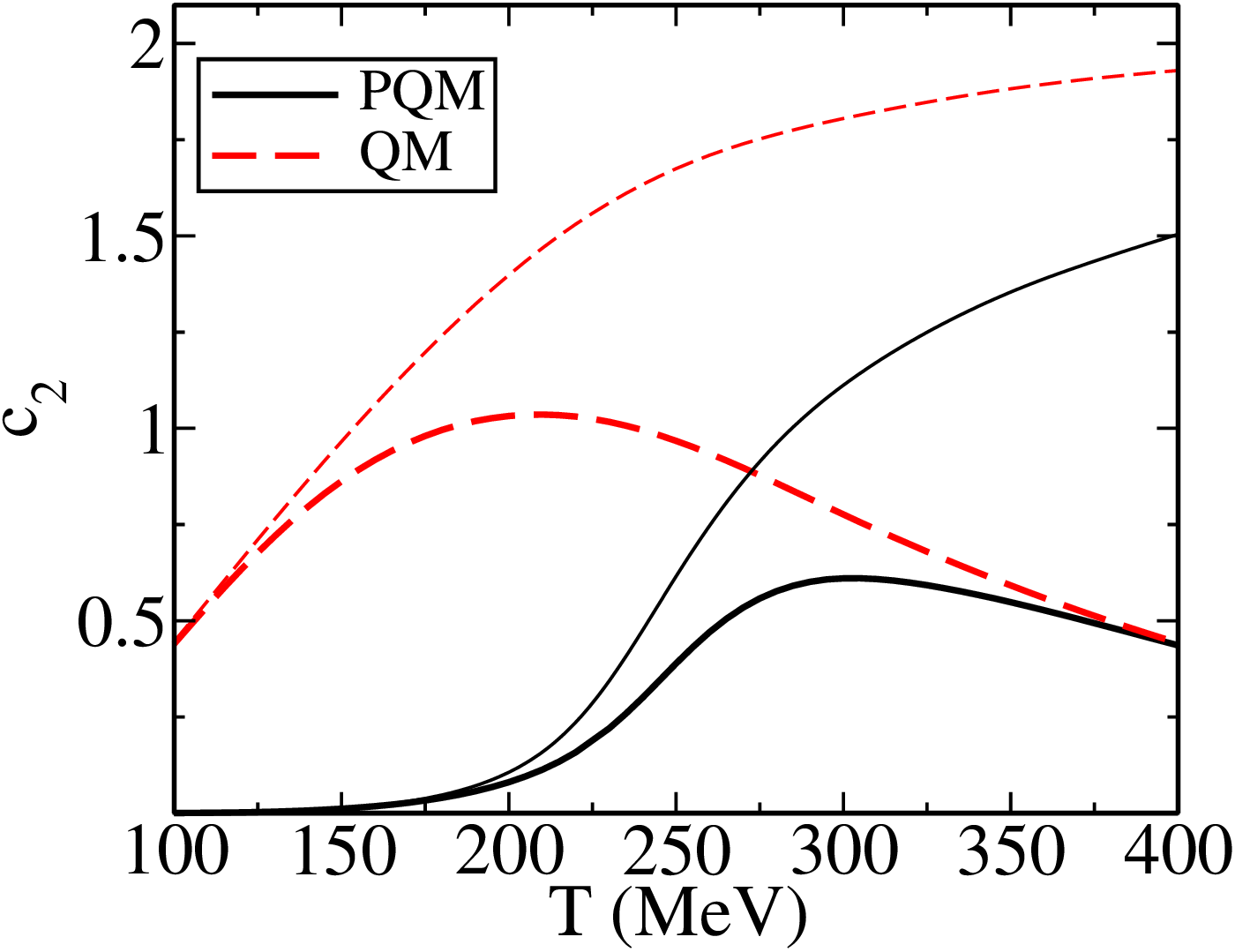}
\caption{
The coefficient $c_2$ as a function of temperature at zero baryon chemical potential
for the PQM and QM models
in the mean-field approximation (left panel) and
in the FRG approach (right panel). The thin lines in the rrightpanel indicate
the FRG results after inclusion of the high-momentum flow (see text for
details). }
\label{fig:c2}
\end{figure*}

\begin{figure*}
\includegraphics*[width=7.25cm]{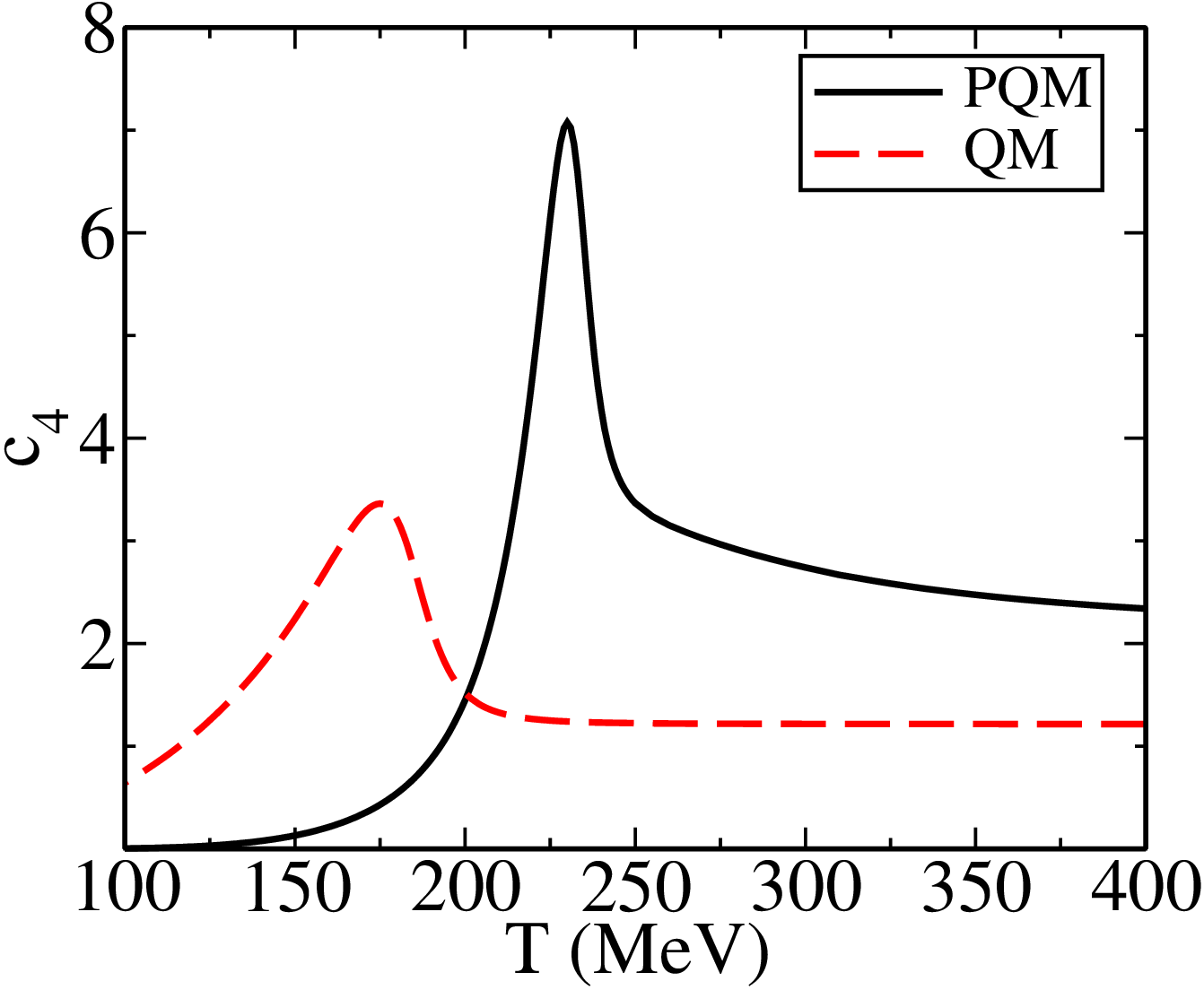}
\includegraphics*[width=7.5cm]{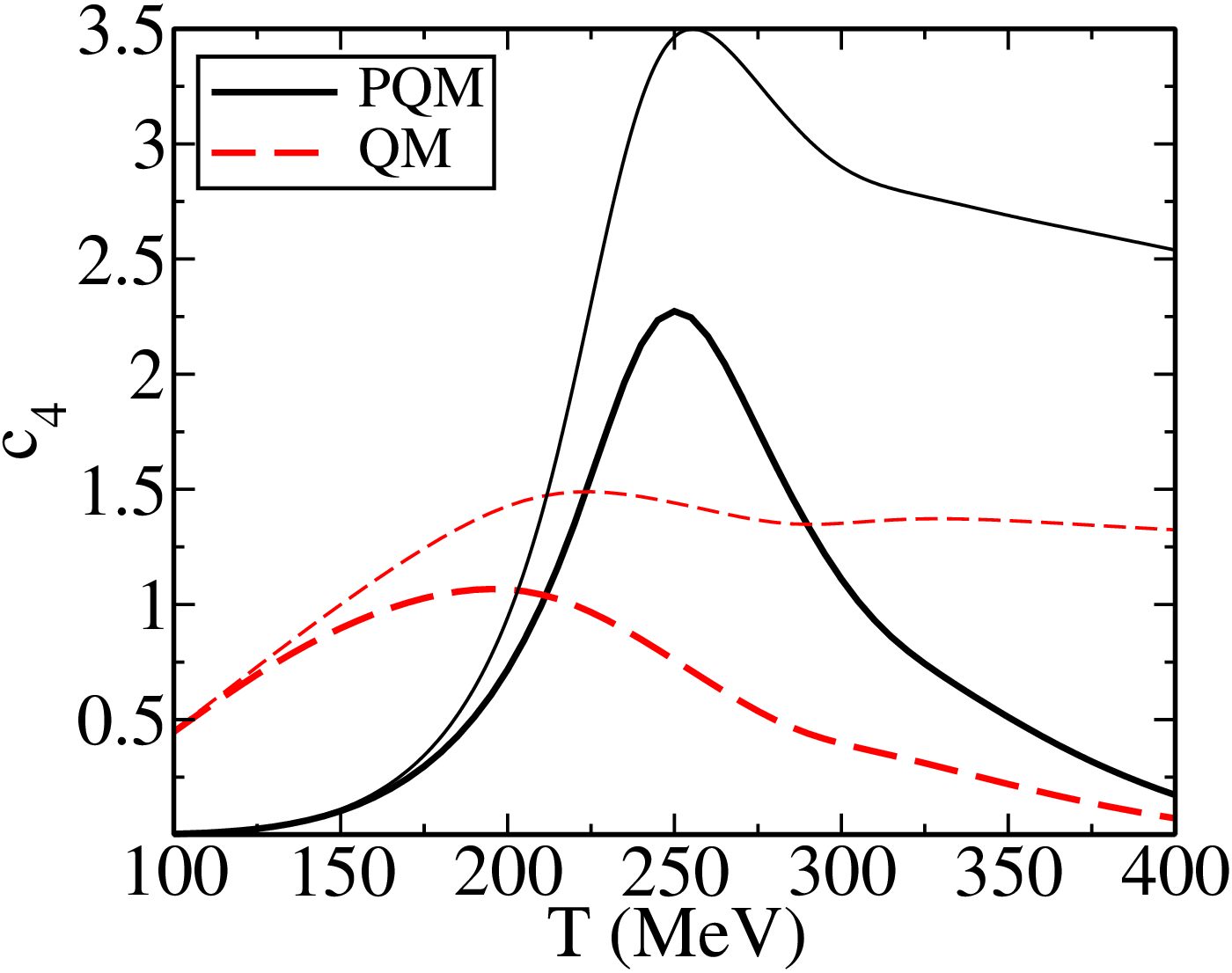}
\caption{
The coefficient $c_4$ as a function of temperature at zero baryon chemical potential
for the PQM and QM models in the mean-field approximation (left panel) and
in the FRG approach (right panel).
As in Fig.~\ref{fig:c2}, the thin lines indicate the
FRG results after inclusion of the high-momentum flow (see text for details).
}
\label{fig:c4}
\end{figure*}

\subsection{Quark number fluctuations}
To explore the  influence of non-zero baryon chemical potential on
the thermodynamics we employ the Taylor-expansion in $\mu/T$ around $\mu=0$
applied in LGT~\cite{LGT} and in model
calculations~\cite{Ratti:2006wg,Ghosh:2006qh}.
In particular, we obtain the expansion of the thermodynamic pressure
\begin{equation}\label{eq:pressure}
\frac{p\,(T,\mu)}{T^4}=\sum_{n=0}^{\infty}\frac{1}{n!}c_n(T)\left(\frac{\mu}{T} \right)^n,
\end{equation}
where
\begin{equation}
\left. c_n(T)=\frac{\del^n[p\,(T,\mu)/T^4]}{\del(\mu/T)^n}\right\vert_{\mu=0}.
\end{equation}

The expansion coefficients $c_n(T)$ are generalized susceptibilities that
characterize the fluctuations of the net quark number $\delta N_q = N_q -
\langle N_q\rangle$ at
vanishing chemical potential~\cite{Allton:2003vx,Ejiri:2005wq,F1}.
In particular, the first two non-vanishing derivatives,  $c_2$
and  $c_4$, are the second and fourth order cumulants:
\begin{eqnarray}
c_2 &=& {{\chi_q}\over {T^2}}=\langle(\delta N_q)^2\rangle,
\nonumber \\
c_4 &=& \langle(\delta N_q)^4\rangle-3\langle(\delta N_q)^2\rangle^2; \label{fluctuations}
\end{eqnarray}
where $\chi_q$ is the regular quark number susceptibility.

%%%%%%

The temperature dependence of the coefficients $c_2$ and $c_4$,  obtained in the
mean-field approximation and in the FRG approach,
is shown in Figs.~\ref{fig:c2} and~\ref{fig:c4}.
The coefficient  $c_2$ increases monotonously with temperature.
In  the mean-field approximation,  $c_2$  increases rapidly in  the critical
region and approaches the ideal gas limit at high temperatures. The
FRG results show a rather different  behavior at high temperatures. The
corresponding susceptibility is strongly suppressed above the chiral transition,
due to the ultraviolet cutoff. The contribution of momenta
beyond the cutoff to the thermodynamics is missing.
In order to obtain the correct high-temperature behavior of
the susceptibilities and other observables, one may supplement the
FRG with the perturbative contribution of the high-momentum states.
A simple model for implementing this correction was proposed in Ref.
\cite{Braun:2003ii}, where the corresponding contribution ($k > \Lambda$) to the
flow is approximated by that of a
non-interacting gas of quarks and gluons. The flow equation for
the high momentum contribution to the QM model then reads
\begin{eqnarray}\label{eq:qcdflow0}
\del_k \Omega_k^{\Lambda}(T,\mu)&=&\frac{k^3}{12\pi^2}\Big\{
2(N_c^2-1)\left[1+2n_B(k;T)\right]
\nonumber \\
&& \hspace*{-2cm}- 4 N_c N_f \Big[ 1-
n_F(k;T,\mu)-n_F(k;T,-\mu)\Big]\Big\},
\end{eqnarray}
where the dynamical quark mass is neglected, i.e. we set $E_q=k$.
For the PQM model we generalize this procedure, by including
the interaction of quarks with the Polyakov loop
\begin{eqnarray}\label{eq:qcdflow}
\del_k \Omega_k^{\Lambda}(T,\mu)&=&-\frac{N_c N_f k^3}{3\pi^2}
\\
&& \hspace*{-1cm} \Big[ 1-
N(\ell,\ell^*;T,\mu)-\bar{N}(\ell,\ell^*;T,\mu)\Big].\nonumber
\end{eqnarray}
Since the effective gluon potential, ${\cal U}(\ell, \ell^*)$, is
fitted to reproduce the
Stefan Boltzmann limit at high temperatures, the explicit gluon contribution is
omitted for consistency.

We thus proceed as follows: Eq.~(\ref{eq:qcdflow0}) or
Eq.~(\ref{eq:qcdflow}) is integrated from $k=\infty$ to
$k=\Lambda$, where we switch to the QM or PQM flow
equation, Eq.~(\ref{eq:frg_flow}).
The divergent terms
in the high-momentum flow equations (\ref{eq:qcdflow0}-\ref{eq:qcdflow}) are
independent of mesonic and gluonic fields, and of temperature and chemical
potential. Consequently, they can be absorbed in an unobservable constant shift
of the thermodynamic potential.

In the right panels of Figs.~\ref{fig:c2} and \ref{fig:c4}
we show the results for $c_2$ and $c_4$ in both the QM and PQM models. We find
a smooth transition between the high and low energy regimes of the theory.
The Stefan Boltzmann limit, $c_2=2$, is reproduced at high temperatures.

% Fig 4
\begin{figure*}
\includegraphics*[width=7.5cm]{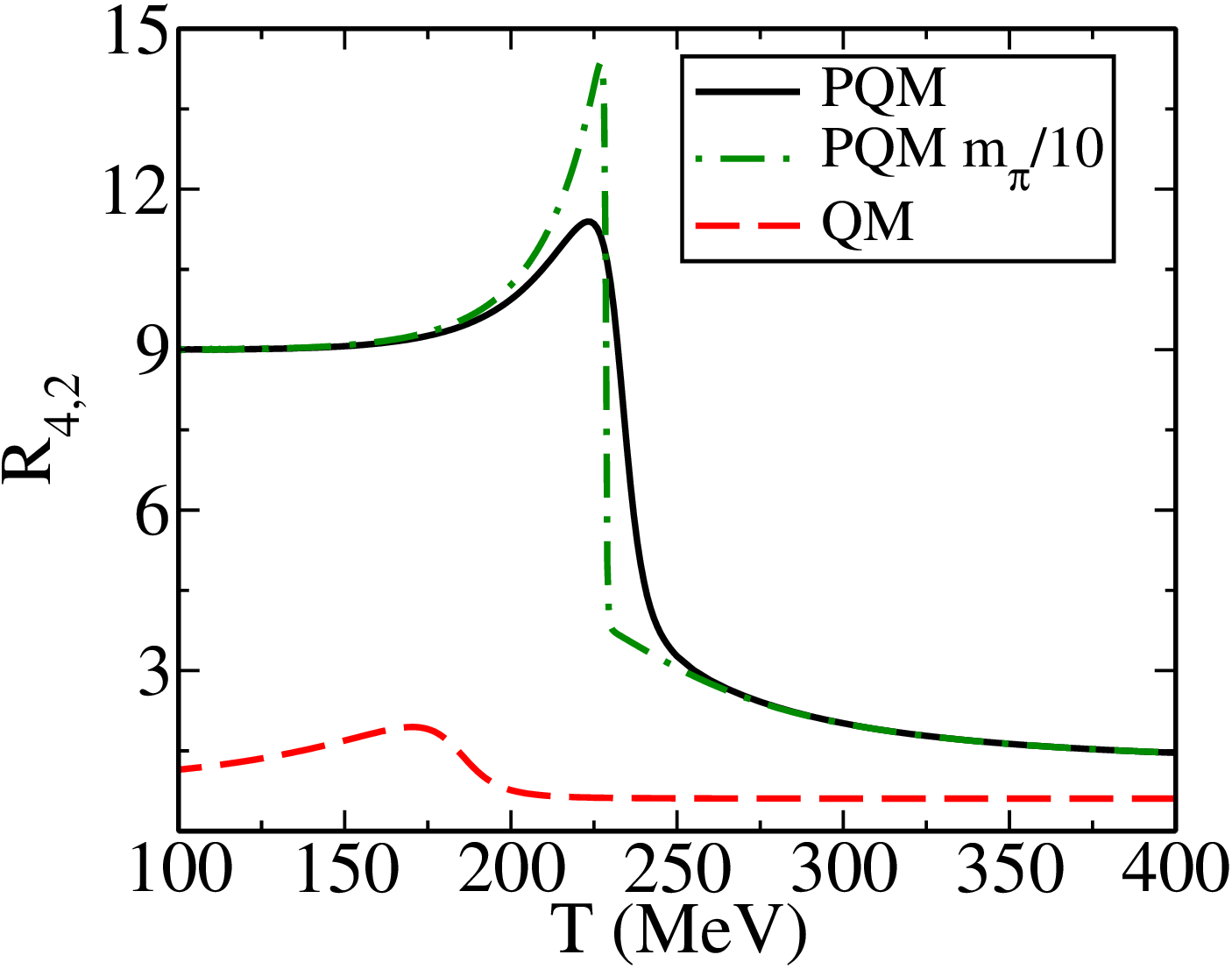}
\includegraphics*[width=7.5cm]{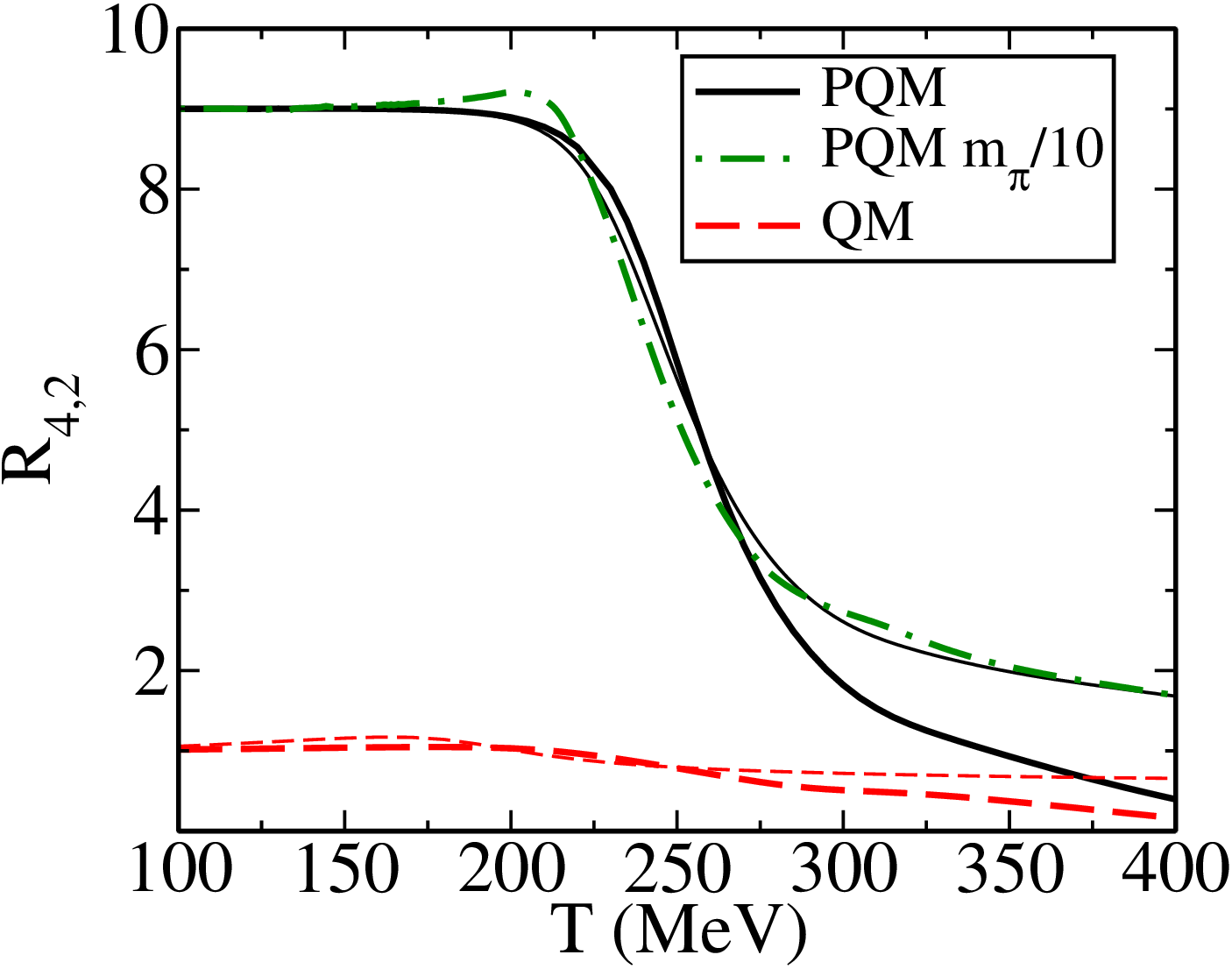}
\caption{
The kurtosis $R_{4,2}$ as a function of temperature at zero baryon chemical potential
for the PQM and QM models  in the mean-field approximation (left panel) and
in the FRG approach (right panel).  The thin lines indicate the
FRG results after inclusion of the perturbative high-momentum modes
(see text for details).
}
\label{fig:r}
\end{figure*}

In contrast to $c_2$,  the coefficient  $c_4$  shows a non-monotonic behavior
near the chiral pseudocritical
temperature.  As shown in Fig.~\ref{fig:c4}, $c_4$ exhibits a pronounced peak at
the pseudocritical temperature.
The shape of the peak is changed considerably when
mesonic fluctuations are included; its
maximum value is reduced while the width is enhanced. We note that the
peak value obtained in the mean-field approximation is strongly dependent on
the cutoff $\Lambda_{MF}$, used to  regularize
the divergent vacuum contribution. A smaller value of the cutoff yields a
larger maximum value of $c_4$. The significance of the fermion
vacuum contribution was recently emphasized in Refs.~\cite{Nakano:2009ps,MFonVT}.

The general trends of the mean-field calculation can be inferred
from Landau's theory of critical phenomena, where
the thermodynamic potential is assumed to be a polynomial in the order parameter
$\sigma$
\begin{equation}
\Omega(T,\mu;\sigma) = \Omega_{bg}(T,\mu) + \frac12 a(T,\mu) \sigma^2 + \frac14
b \sigma^4 +   c \sigma.
\label{LG}
\end{equation}
Here  $\Omega_{bg}(T,\mu)$ is a background contribution, independent of
$\sigma$, and
$c$ is a symmetry breaking term. The coefficient $a$
is generally assumed to be a linear function of the temperature, $a=A\cdot(T-T_c)$,
where $T_c$ is the critical temperature of the second-order phase transition
for $c=0$. An extension to small, non-zero values of the chemical potential is
achieved by assuming
\begin{equation}
a(T,\mu) = A\cdot(T-T_c)  + B\mu^2,
\label{a}
\end{equation}
where both coefficients $A$ and  $B$ are positive. In general the effective quartic coupling constant, $b>0$,
also depends on temperature and chemical potential. However this
dependence is irrelevant for  $T\simeq T_c$ and $\mu\simeq 0$.  In the chiral
limit, $c=0$, the $c_2$ and $c_4$ coefficients are easily obtained from
Eq.~(\ref{LG}):
\begin{eqnarray}
c_2 &=& c_{2 bg} + \frac{A B}{b T} \frac{T-T_c}{T} \theta(T_c-T), \\
c_4 &=& c_{4 bg} + \frac{6 B^2 }{b} \theta(T_c-T).
\label{c2c4LG}
\end{eqnarray}
Thus,  $c_2$ is not differentiable at the critical
point, while $c_4$ exhibits a discontinuity. The background parts, $c_{2 bg}$
and $c_{4 bg}$, are
smooth functions of temperature and do not change the critical behavior. For a
finite pion mass ($c\neq0$), the transition is of the cross over type and the
sharp structures in $c_2$ and $c_4$ are smoothened. Consequently, in the QM and
PQM models, the peak structure appearing in $c_4$ is due to the decreasing quark
mass, and thus closely connected to the restoration of chiral symmetry. By
contrast, in the resonance
gas model, the peak in $c_4$ is due to the increasing contribution of
higher resonances.

In general, the singular part of thermodynamic potential is
in chiral limit controlled by the critical exponents of the tree-dimensional
O(4) symmetric spin model.
Consequently, at zero chemical potential:
\begin{eqnarray}
\Omega_{sing} &\sim& (T-T_c)^{2-\alpha}, \\
c_{2 sing} &\sim& (T-T_c)^{1-\alpha}, \\
c_{4 sing} &\sim& (T-T_c)^{-\alpha}.
\label{scaling}
\end{eqnarray}
Critical fluctuations renormalize the exponent $\alpha$ from its mean-field
value $\alpha_{MF}=0$ to $\alpha=-0.21$. Consequently, fluctuations lead to a
weakening of the singularity in the chiral limit. This is reflected in a
smoothening of the temperature dependence for finite pion mass, as seen in
Figs.~\ref{fig:c2} and \ref{fig:c4}. We note that in QCD the singularity in
$c_2$ and $c_4$ may be obscured by resonances, which presumably dominate the
thermodynamics of the low-temperature phase.

A non-vanishing background gluon  field leads to qualitative changes of the
quark fluctuations. In the low-temperature phase,  there is a
strong suppression of the quark
fluctuations due to their coupling to the Polyakov loop (see Figs.~\ref{fig:c2}
and~\ref{fig:c4}). This is
because single and double quark
modes are suppressed. when the expectation value of the Polyakov loop, $\ell$
and $\ell^\star$, is small (see Eqs. (\ref{eq:frg_flow},\ref{n1}) and
(\ref{Omega_MF_q},\ref{g})).

Lattice studies of QCD at finite temperature and density as
well as chiral
model calculations show, that the ratio (kurtosis)
\begin{equation}\label{eq:ratio_c42}
 R_{4,2}=\frac{c_4}{c_2}
\end{equation}
is an useful probe of the deconfinement and chiral phase
transitions~\cite{Ejiri:2005wq,F1,kurtosis}. In the
high and low temperature regimes, the kurtosis reflects the
net quark content of
the dominant baryon number carrying effective degrees of freedom
~\cite{Ejiri:2005wq,kurtosis}.

In Fig. \ref{fig:r} we show the ratio $R_{4,2}$  as a function of temperature.
Both in the FRG approach and in the mean-field approximation for the PQM model,
the kurtosis drops from $R_{4,2}\simeq 9$ to $R_{4,2} < 1$ in the transition
region, as expected due to the change in quark
content of the baryon carrying effective degrees of
freedom~\cite{Ejiri:2005wq}. At low temperatures
the effective three quark states dominate, while at high temperatures single
quarks prevail. In the mean-field approximation (see the left  panel of
Fig.~\ref{fig:r}), the kurtosis exhibits a well defined peak at the transition
temperature. The height of the peak depends
on the value of the pion mass~\cite{kurtosis,karsch}. This dependence is
illustrated in Fig.~\ref{fig:r} both in the mean-field approximation and in the
FRG approach. The inclusion of mesonic fluctuations weakens the
dependence on the pion mass. Thus, for a physical value of $m_\pi$, the kurtosis
decreases monotonously with temperature in the FRG approach.
A peak in $R_{4,2}$ appears only for a pion mass below its physical
value. In the mean-field approximation, the dependence on the pion mass is
more pronounced, but still not as dramatic as found in Ref.~\cite{kurtosis}.
This is due to the different treatments of
the divergent fermion vacuum contribution. In Ref.~\cite{kurtosis} the vacuum
term was subtracted at $T=0$ and its dependence on temperature was neglected.
Here we regularize the vacuum term with an ultraviolet cut off and retain its
temperature dependence, which is due to the in-medium quark mass.
The role of the vacuum term in mean-field calculations
and its influence on the chiral phase transition
are explored in detail in Ref.~\cite{MFonVT}.

In effective quark models where the Polyakov loop is neglected, the
thermodynamics is governed by one-quark states at all temperatures.
Consequently, in
the QM model the low-temperature limit of the kurtosis is $R_{4,2}=1$ rather
than 9. Consequently, in such models, the interaction of quarks with the
background Polyakov loop is essential for reproducing the low-temperature
behavior of  $R_{4,2}$  found in LGT
calculations.

%%%%%%%%%%%%%%%%%%%%%%%%%%%%%%%%%%%%%%%%%%%%%%%%%%%%%%%%%%%%%%%
\section{Summary and Conclusions}\label{sec:concl}
We have discussed  thermodynamic properties of the Polyakov loop extended quark--meson effective
chiral model   (PQM), including fluctuations within the functional
renormalization group method
(FRG). A truncation of the PQM model was introduced, which allowed us  to
extend  previous
renormalization group studies by introducing the coupling of fermions to the
Polyakov loop. We have thus
formulated and solved the  flow equation for the scale dependent thermodynamic potential at finite
temperature and density in the presence of a background gluonic field.

In our studies of the thermodynamics we have  discussed the role of
fluctuations on the critical
properties of the model. 
%We have shown that the non-perturbative dynamics
%introduced in FRG approach
%modifies   the phase diagram, leading to   a shift in the position of the
%critical end point (CEP).
{ The fluctuations of the net quark number density and
the ratio of the fourth to second order cumulants, the kurtosis, were
analyzed.

Furthermore, a comparison of the FRG approach
with and without a gluon background field  was  performed.} The influence of
ultraviolet cutoff
effects on the thermodynamics within the FRG approach was also discussed. We
have shown that the FRG extended
quark--meson model preserves  the basic properties of the kurtosis, obtained in
LGT calculations only if
the Polyakov loop is included in the flow equation. Thus, the extension of the
FRG method proposed
here, { accounting for the} coupling of fermions to background gluon
fields, is
of particular relevance for effective descriptions of QCD thermodynamics near
the phase transition in terms of the quark--meson model.

%%%%%%%%%%%%%%%%%%%%%%%%%%%%%%%%%%%%%%%%%%%%

\section*{Acknowledgment}
We acknowledge fruitful  discussions with E. Nakano and B.J. Schaefer. 
V.S. acknowledges stimulating discussions with K. Fukushima.
K.R. acknowledges partial support from the Polish Ministry of Science (MEN)  and the Alexander von Humboldt Foundation (AvH). B.S. gratefully acknowledges financial
support from the Helmholtz Research School on Quark Matter Studies.
%%%%%%%%%%%%%%%%%%%%%%%%%%%%%%%%%%%%%%%%%%%%%%%%

\end{document}